\documentclass[12pt]{iopart}

\usepackage{graphicx}
\usepackage{multicol}
\usepackage{cite}
\begin{document}

\title[Intrinsic decoherence and purity in a BEC]{Intrinsic decoherence and purity in a Bose quantum fluid in a triple well potential}

\author{A Camacho-Guardian and R Paredes}

\address{ Instituto de F\'{\i}sica, Universidad
Nacional Aut\'onoma de M\'exico, Apdo. Postal 20-364, M\'exico D.
F. 01000, M\'exico.}
\ead{rosario@fisica.unam.mx}
\begin{abstract}
We consider a quantum Bose fluid confined in a triple well potential in 1D within the exact $N$-body Bose-Hubbard model to investigate the phenomena of intrinsic decoherence and loss of purity. Our study is done by following the time evolution of one-body properties in an $N$-particle closed environment. We do an exhaustive exploration of initial conditions to characterize these phenomena. Here we illustrate our main findings with a set of relevant Fock and SU(3) coherent states. Our study shows that signatures of stationarity and maximal mixing are a direct consequence of the inter-particle interactions in the closed system and become evident as the number of particles is increased. This fact is confirmed by quantifying the deviations from stationarity by means of a matrix norm. 
\end{abstract}

\maketitle
\begin{multicols}{2}
\section{Introduction}
The capability to control and manipulate the interatomic interactions as well as the confinement potential where they move, have made ultracold atomic gases excellent and clean experimental systems where predictions and phenomena belonging to the field of condensed matter can and have been reenacted \cite{Bartenstein, Zwierlein, Zwierlein06, Gaebler,Weinmann, Lahaye,Clement,Lye}. This has yielded the opportunity to probe into questions that may not be possible to do directly in the latter systems. In this regard, large ensembles of bosonic atoms confined in optical lattices have been thought as promising candidates in the realization of controllable macroscopic coherent states to pursue quantum computing objectives. This is so because, since the first realization of an optical lattice, signatures of macroscopic coherence were observed \cite{Andrews, Cataliotti, Bloch}. 

Most of the phenomena alluded above are based on the existence of macroscopic quantum coherence. The opposite, decoherence, is a process in which the phase coherence or phase interference is destroyed. Due to unavoidable presence of interatomic interactions, even when they can be externally handled and thus diminished, there exists a characteristic time where coherence diminishes \cite{Sinatra}. Since typical measurements carried out in optical lattices are related with oscillations in particle population across lattice wells, these systems can be thought as suitable devices to be used as time coherence tracers. In particular, Bose gases confined in arrays of few wells may be considered as ideal realizations to pursue such aims. A condensate confined in a double-well potential, the so called bosonic Josephson junction\cite{Albiez}, represents the simplest system where decoherence can be investigated \cite{Bar-Gill,CaballeroD}. 

Theoretical studies of optical lattices have been accomplished within mean field (MF) and Bose-Hubbard (BH) schemes\cite{Jaksch}. As it is well known both descriptions predict the opposite regimes of Josephson oscillations (JO) and self-trapping (ST) as a function of the ratio between inter-particle interactions and hopping strengths. However, the most relevant discrepancy among them is that, while the transport of particles across the lattice in MF approximation shows always coherent oscillations, except for exceptional values of the initial conditions associated to fixed points, in BH description, typically, the dynamics shows collapses to quasi-stationary states with revivals in the particle population. These collapses are caused by the unavoidable presence of the inter-particle interactions, and are observed in actual experiments \cite{Bar-Gill}. Even though interatomic interactions can be externally changed by means of Feshbach resonances, it is almost impossible to completely neutralize them. It is this behavior of collapses and revivals one of our main interests here. That is, we will show, as pointed earlier \cite{Caballero}, that as the number of particles $N$ is increased, for fixed interactions, the time spent in the revivals becomes much smaller than the time spent in the collapsed intervals, in such a way that mostly the system is at a states that become effectively stationary. We call these states statistically stationary and the decay to those states as ``intrinsic" decoherence.

In this work we focus in the study of these aspects for bosons confined in triple wells potentials in 1D (TWP1D) within the Bose-Hubbard approach. The dynamics and stationary properties of condensates in TWP1D have been widely investigated within the mean field or Gross-Pitaevskii approach; the latter studies encompassing transport characterization of chaos \cite{Liu, Mossmann,Franzosi, Franzosi03, Buonsante03, Hennig, Nemoto}, irreversible transport between wells \cite{Graefe,Nesterenko,Rab}, dynamics of collective excitations \cite{Viscondi} and tunneling inhibition as prospects for transistor-like quantum engineering \cite{Stickney,Schlagheck} among others. In the context of BH or full quantum description we can mention the analysis of stationary states and tunneling dynamics for few ($N= 3, 4, 5, 7$) \cite{Wright, Lushuai,Jason} and large ($N= 500$) ensembles of bosons \cite{Paredes}.

As stated, the main interest of this manuscript is to characterize the processes of intrinsic decoherence and loss of purity in a Bose condensate confined in a TW1D. We emphasize that this decoherence is termed ``intrinsic" to distinguish it from the relaxation process that takes place when a quantum mechanical system weakly interacts with its environment. However, from another perspective, the difference may not be as strong as it may appear, since the environment in this case may be represented by the $N$-particle states while the ``system" is associated with few-body properties. As discussed in Ref.\cite{Milburn}, intrinsic decoherence can be addressed from two different approaches. By means of modifications to the unitary evolution of Schr\"odinger evolution or by strict quantum mechanical procedure plus statistical mechanical arguments. Here, we adopt the second one, to study the entitled intrinsic decoherence in an interacting Bose-Einstein condensate in a symmetric triple well potential in 1D. As stated in Ref. \cite{CaballeroD}, a synonymous of intrinsic decoherence is the observation of stationary states; that may or may not be thermodynamic equilibrium states. Such stationary states are characterized by having their few-body properties no longer evolving in time. The attainment of such a stationary state can be monitored by following the evolution in time of the matrix elements of the few-body reduced density matrix $\rho^R$ which, once in the stationary state, those matrix elements become constant. Such a constancy in time has been identified as a signature of decoherence since it is always possible to find a proper basis, the so-called  {\it preferred basis} \cite{Joos, Zurek}, in which the off-diagonal matrix elements of the reduced density matrix become zero. Another quantity recorded from the dynamics within the full quantum scheme, that allows to exhibit the influence of inter-particle interactions in a closed isolated system, is the purity $\mathcal{P}$ of the state. This quantity defined as $\mathcal {P} = \mathrm {Tr} \rho^2$, with $\rho$ the density matrix of the state \cite{Viscondi1,Chianca, Somma}, measures the degree of decoherence or mixing in a system. A totally coherent state has $\mathcal {P}=1$ indicating that the system state can be specified by a vector state $|\Psi \rangle$ such that $\rho =| \Psi \rangle \langle \Psi |$. As it is known, the specification of such a state requires to perform a complete set of measurements consistent with the system degrees of freedom. In analogy to the intrinsic decoherence detected from $\rho^R$, here we shall consider the purity of one-body states. Namely, in this work we study the evolution in time of the one-body reduced density matrix, and the purity $\mathcal{P}$, to demonstrate how an interacting $N$-particle system described through a Bose-Hubbard Hamiltonian exhibits both intrinsic decoherence and loss of purity. 
    
This paper is organized in four sections. In section II we present the model that describes the system, namely the BH Hamiltonian, and summarize in a phase diagram all the stationary states as a function of the number of particles and interaction strength.  In section III we concentrate in studying certain aspects of the full quantum dynamics that emerge as the system size is increased, namely the intrinsic decoherence and the purity. By using relevant sets of $N$-particle systems we illustrate in this section the existence of a stationary state and the purity in body properties. Our calculations are performed for $N=10, 40, 80, 120$ and 150. To quantify the deviations from the stationary mean value as a function of $N$ we use the Frobenius norm. Finally, in section IV a summary of our findings is presented. 

\section{A three well potential in the one-level picture, stationary eigenstates}

The system under study consists of a Bose quantum fluid confined in a symmetric triple well potential in 1D (TW1D). As stated in Section I, stationary and dynamical properties of this system has been extensively studied in the literature within BH and MF schemes \cite{Hennig,Nemoto,Franzosi, Franzosi03,Buonsante03,Mossmann,Liu,Graefe,Rab,Nesterenko, Viscondi,Schlagheck,Stickney,Wright, Lushuai,Jason,Paredes, Paredes09}. In this section we revisit the analysis of the stationary states for the TW1D. The aim of such a study is to compare the stationary states of BH model obtained from direct diagonalization, with those predicted from the $N$-body dynamical evolution. The study of stationary states from dynamical evolution will be performed through the analysis of decoherence and loss of purity.  

Within the full quantum $N-$body scenario, except for constant factors that scale with the number of particles, the Hamiltonian that describes the system in the one level-picture, that is, considering the three lowest single particle states and its corresponding Wannier functions, is the BH model \cite{Jaksch, Paredes}

\begin{eqnarray}\fl \nonumber
H=-J(b_1^{\dagger}b_{2}+b_2^{\dagger}b_{1}+b_2^{\dagger}b_{3}+b_3^{\dagger}b_{2})\\+U_0\sum_{i=1}^3b_i^{\dagger}b_i^{\dagger}b_{i}b_{i},
\label{h3}
\end{eqnarray}
where the operators $b^\dagger_i$ and $b_i$  create and annihilate particles in the site $i$ and satisfy the usual commutation rules for bosons $[b_i,b^\dagger_j]= \delta_{ij}$. The interaction strength $U_0$ written in terms of the $s$-wave scattering length $a$ , $U_0=(2\pi \hbar^2 a/m) \int \psi_i^4(x)dx$, with $m$ the atomic mass of the considered specie and $\psi_i(x)$ the localized Wannier functions, scales the contact interactions among pairs of particles.  For a given number of particles $N$, the size of the Hilbert space is $\Omega = (N+1)(N+2)/2$ and therefore, the size of Hamiltonian, given by Eq. (\ref{h3}), scales as $ \sim N^2 \times N^2$. The information of the energies of the Hamiltonian stationary states can be encoded in a phase diagram as a function of the parameter $\Lambda=U_0N/J$ that measures the ratio among the interaction $U_0$ and the tunneling amplitude $J$ \footnote{The coefficient of the tunneling amplitude $J$ is related with the inverse of the oscillation period $\tau$ of a single particle in the symmetric triple well potential as $J= h/\tau$. For our numerical calculations we assume $J=1$ in dimensionless units. In an experiment this can be done by adjusting the depth of the optical potential, that is, the laser intensity.}.  For our calculations we consider $N = 150$ \footnote{The maximum value of $N$ for which we can perform calculations in a reasonable time is $\sim 200$ but taking $N=150$ leads quantitatively to the same features than $N=200$. For practical reasons we restrict ourselves to consider $N=150$ for dynamical calculations.}. Early\cite{Wright} and recent\cite{Jason} works report calculations for $3<N<90$ for symmetric arrays of trimers in open and closed configurations.

\begin{figure*}[htbp]
\centering
\includegraphics[width=3.0in]{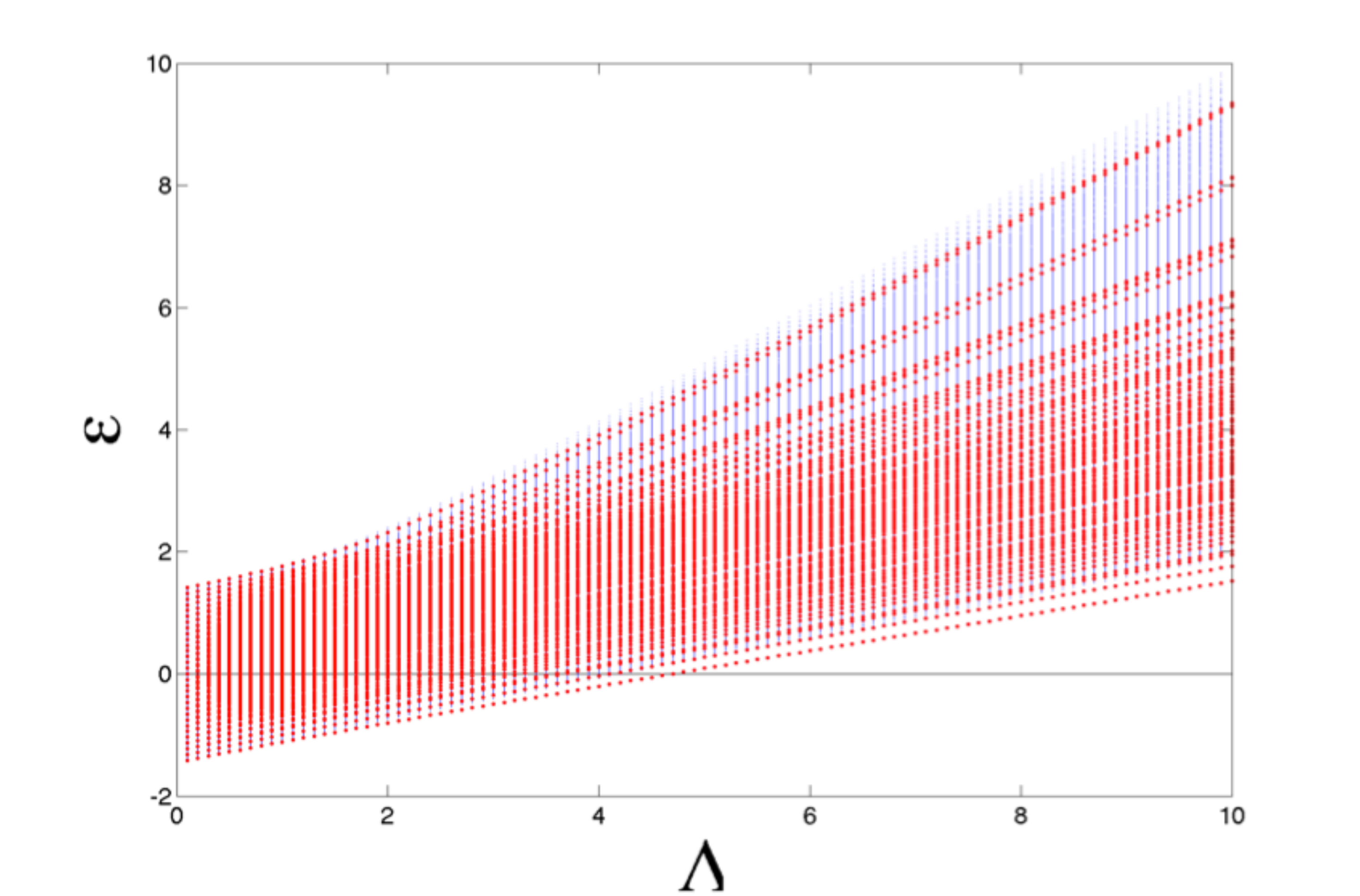} 
\caption{(Color online) Exact energy eigenvalues $\varepsilon=E_n/NJ$ of the $N-$particle Hamiltonian (\ref{h3}) as a function of  $\Lambda =U_0N/J$. Red and blue dots  correspond to $N=15$ and 150 respectively.}
\label{Fig1}
\end{figure*}

In Fig.\ref{Fig1} we plot the exact energy eigenvalues of Hamiltonian (\ref{h3}) as a function of $\Lambda$, for $N=15$ and $N=150$, and for values of $\Lambda$ ranging from $0.0 \le \Lambda \le 10.0$, in steps of $0.1$.  All the eigenvalues associated with the ground state lie on a straight line with constant slope, while the highest eigenvalues lie on a curve that changes its slope at $\Lambda \approx 1.5$. As it is known \cite{Franzosi}, such sets of minimum and maximum eigenenergies correspond to states in which particles oscillate coherently or remain trapped respectively, the so called JO and ST regimes. Since this transition has been studied and characterized in the BH context for the TW1D, here we just want to emphasize the dependence (dense or sparse) of the energy spectrum on $N$. We shall name these energy eigenstates as the truly stationary since their evolution remains unchanged by the full unitary propagator $U= \exp [-i Ht/\hbar]$. 

Regarding the MF scenario, it is well known that even though its intrinsic nonintegrability, a BEC confined in a triple well potential represents the simplest nonintegrable system that admits, an exhaustive numerical description. For comparison purposes with their BH counterpart, we determine the stationary states (see Fig.\ref{FigA}). The straightforward procedure to obtain those stationary states is summarized in the Appendix.

\section{Quantum dynamics} 

In the previous section we determined the stationary eigenstates of the symmetric triple well potential in the BH scheme. We now turn our attention to the study of certain dynamical aspects of one-body properties with the interest of exhibiting  phenomena that appears only within the $N$-particle full-quantum description. First, we shall concentrate in describing the so-called phenomenon of intrinsic decoherence and then we analyze the loss of purity in the closed $N$-particle system. These phenomena can emerge within the full quantum scheme only and its manifestation becomes more evident as the number of particles is increased.

\subsection {Stationary states and intrinsic decoherence}
\label{SS-ID}

The phenomenon of intrinsic decoherence becomes evident in large $N$-body closed systems when they are evolved in time and a statistically stationary state is observed. Such a state can be registered only through the behavior of few body properties, when they exhibit an oscillatory behavior that decays to a constant value that no longer changes during a period of time large enough compared with a characteristic system time scale. Then, after remaining with that constant value during a time interval, the dynamics show again an oscillatory behavior for another time interval \footnote{According with the Poincar\'e theorem, certain systems will, after a sufficiently long but finite time, return to a state very close to the initial one, the so called recurrences.}, before decaying to the same constant value. The elapsed time to reach such a constant value and the time during which the value of the property under study remains constant are called the relaxation $\tau_{rel}$ and the recurrence $\tau_{rec}$ times respectively. In general, the values of both $\tau_{rel}$ and $\tau_{rec}$ depends strongly on the system size and on the initial state from which the system is evolved. Ultracold bosons confined in optical lattices are ideal systems where these assertions, for macroscopic conglomerates, can be reproduced. The occurrence of a state very close to the initial one, starting from any arbitrary initial condition, is guaranteed because of the fact that within the BH model the Hilbert space has a finite size. For $N$ particles confined in a 1D lattice composed of $n$ wells it scales as $\left(\begin {array}{c} N+n-1 \\ N \end{array} \right)$, as consequence of considering that the Hamiltonian that describes the system takes into account  the single particle-modes of the first band only. It is worth to mention that as the number of wells $n$ and particles $N$ is increased, the numerical analysis becomes impracticable. Regarding the dependence of $\tau_{rel}$ and $\tau_{rec}$ on the system size, it has been shown for arrays of two wells (using a particular initial condition), that those times scale in such a way that in the limit of large $N$, $\tau_{rel}/ \tau_{rec} \rightarrow 0$, thus showing the existence of a stationary state, that we label as `statistical" to differentiate it from the stationary eigenstates. Here, we concentrate into characterize the phenomenon of intrinsic decoherence for the TW1D by studying the evolution in time of one body properties. In particular, we shall show that for every initial condition the evolution in time of those properties exhibit on average constant values that no longer evolve in time. We shall also analyze the dependence of the deviations around the mean stationary value on the system size.

The attainment of a state in which the few body properties no longer change, within the fluctuations around a mean value, identifies the statistically stationary states. The way in which is possible to detect such states is by following the evolution in time of the expectation value of few body properties for arbitrary initial states. It is not necessary to argue that if the initial state is chosen to be an $N$-particle energy eigenstate, the unitary evolution will leave it invariant, and thus any few body property will neither evolve from its initial value. The expectation value of any few-body property can be determined from either the $N$-body density matrix or the few-body reduced density matrix $\rho^R$. As mentioned in the previous paragraph,  here we concentrate in the analysis of one particle properties only. Let $O^{(1)}$ be an one body operator of the $N$-particle system. That is $O^{(1)}=\sum_{m=1} ^N O_{m}^{(1)}$  where  $O^{(1)}_m$ represents a single particle operator. The expectation value of such an operator is given by

\begin{equation}\fl
\langle O^{(1)} (t)\rangle= {\rm Tr} \rho^N(t) O^{(1)},
\label{O_1}
\end{equation}
where  $ O^{(1)} = \sum _{i,j} \langle i |  O^{(1)} | j \rangle b_i^\dagger b_j$, being the single particle states represented by $| i\rangle$ with $i =1,2,3$ in the present problem. Written in terms of the one body reduced density matrix,
\begin{equation}\fl
\rho_{ji}^R(t)=\frac{{\rm Tr} (b_j\rho^N(t) b_i^\dagger)}{N},
\end{equation}
 the expectation value of $O^{(1)}$ is
\begin{equation}\fl
\langle O^{(1)} (t)\rangle=  \sum_{i,j} \langle j | O^{(1)} |i \rangle \rho_{ji} ^R.
\end{equation}

To demonstrate that any arbitrary one body property attains a stationary state it suffices to show that, for a given initial condition, every element of the one-body reduced density matrix reaches, within fluctuations around a mean stationary value, a constant value. In addition, to provide one of the signatures that accompanies the manifestation of decoherence, one has to show that in the stationary state the one body reduced density matrix is a diagonal matrix. That is, 

$$
\rho_p^R= \mathcal{U} \rho^R_s \mathcal{U^\dagger} = \left( \begin{array}{ccc}
\rho_{11}^p & 0 & 0\\
0& \rho_{22}^p & 0 \\
0 & 0 & \rho_{33}^p \end{array} \right).
$$
where $\rho_{ii}^p$  $i= 1,2,3$ denote the mean stationary values that reaches the diagonal elements of $\rho^R$ in a particular basis, the so called preferred basis \cite{Joos, Zurek}.  As we shall see, this basis defined in terms of the unitary transformation labeled by the operator $ \mathcal{U}$ can be obtained numerically and depends on the initial condition and on the value of $\Lambda$.
 
The $N$-particle density matrix in Eq. (\ref{O_1}) is given by $\rho^N (t)=  |\phi (t) \rangle  \langle \phi (t) | $, with $|\phi(t) \rangle = e^{-i Ht/\hbar} |\phi(0)\rangle$ where $|\phi(0)\rangle$ is the initial state. Given the eigenstate basis or the number-Fock basis, there exist an infinite number of initial states that can be used to analyze the behavior of the one body reduced density matrix $\rho^R (t)$. We did an exhaustive exploration considering as initial states those that constitutes de Fock basis and we also considered the SU(3) coherent states \cite{Wright}. Our analysis was performed for values of $\Lambda$ below and above the transition from JO to SF regimes. We illustrate the generality of our findings by showing first the evolution in time of $\rho^R (t)$ for {\it i}) initial SU(3) and then, we show the results for  {\it ii}) an initial state constructed as a superposition of the number-Fock states. The SU(3) states are defined as
\begin{equation}\fl
|\Phi\rangle_{{\rm SU(3)}}=\frac{1}{\sqrt{N!}}\left(\sum_{i=1}^{3}\frac{\psi_i}{\sqrt{N}}b_i^{\dagger}\right)^N|0\rangle, 
\label{SU3}
\end{equation}
being $|0\rangle$ the vacuum state and $\psi_i$ a complex quantity describing the bosons at lattice site $i$ through a macroscopic local phase $\phi_i$ and population $|\psi_i|^2$, that is, $\psi_i= \sqrt{Nn_i} e^{i \phi_i}$. The special feature that the SU(3) states have is that the quantum expectation value of any operator, and in particular the number operator, can be related to the solutions of the MF coupled equations \cite{Wright}.  For this reason if $\psi_i$ numbers are chosen to be the stationary solutions of the Hamilton equations for a given value of $\Lambda$ (the fixed points), then the system will remain in a stationary state under the action of time evolution operator $U$. 

\begin{figure*}[!htbp]
\centering
{\includegraphics[width=116mm,height=80mm]{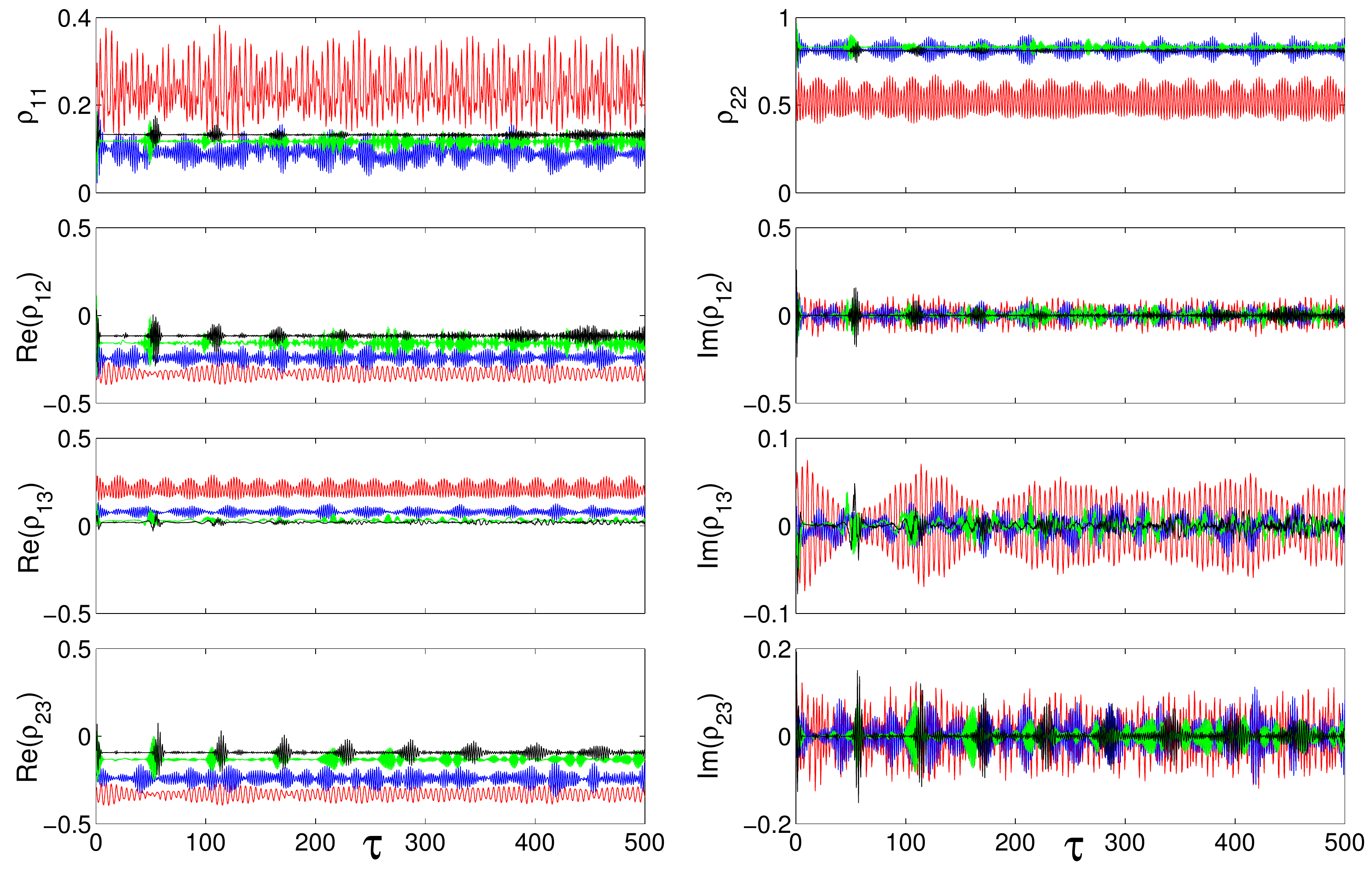}}
{\includegraphics[width=80mm,height=30mm]{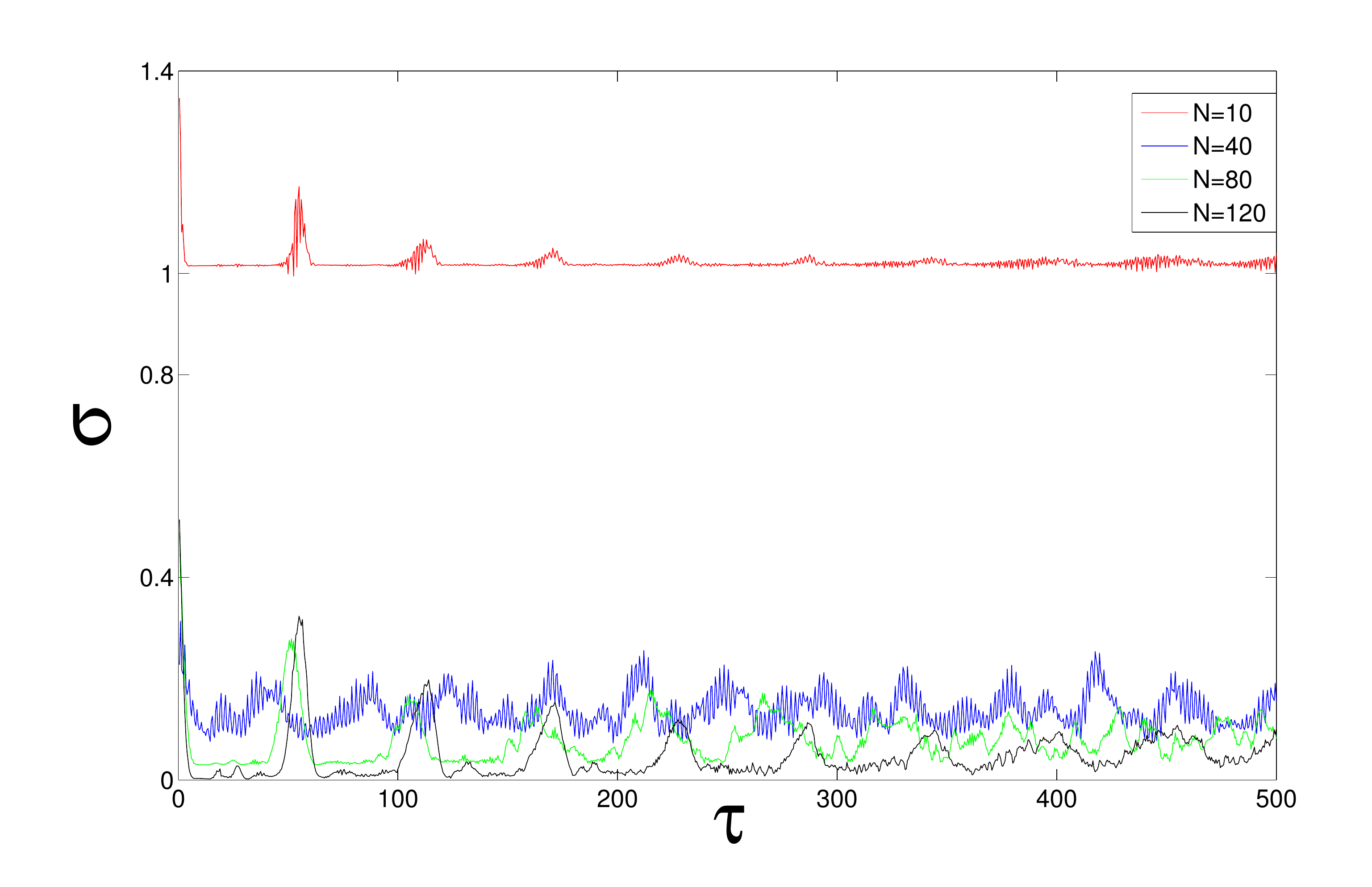}}
\caption{(Color online) Real and imaginary parts of the diagonal and off-diagonal elements of the one-body reduced density matrix as a function of $t$ for $N=10,40,80$ and 120. The initial condition is defined by equation (\ref{SU3}) with $n_1=0.2,$ $n_2=0.7$, $n_3=0.1$ and $\phi_{12}=\phi_{32}=\pi$. $\tau$ is in dimensionless units.}
\label{Fig2}
\end{figure*}

Here and henceforth the time $t$ in each figure has been rescaled to dimensionless units $\tau= Jt/ \hbar$ \cite{J}. In typical experiments the values of $J$ and $U_0$ remain constant \cite{Albiez}. For our numerical calculations we consider  $N= 10, 40, 80$ and 120 and $U_0/J= 0.05$ fixed. The first initial condition that we select to illustrate our results  belongs to the SU(3) coherent states defined by Eq. (\ref{SU3}), specifically we chose $n_1=0.2$, $n_2=0.7$, $n_3=0.1$, $\phi_{21}=\phi_{32}= \pi$, with $\phi_{ij}= \phi_j-\phi_i$.  Such condition correspond to a non symmetric state. In Fig. \ref{Fig2} we plot the real and imaginary components of the elements of the one-body reduced density matrix $\rho_{i j}^R$ for different values of $N$ and a constant interaction strength $U_0$. In the basis considered, that is, the atom number state or Fock basis, the terms $\rho_{11}(t)$ and $\rho_{22}(t)$ give the particle population fraction for each of the wells, while the terms $\rho_{ij}(t), \> i \ne j$ are the so called coherences. As one can see from this figure all the elements of $\rho^R$ show an oscillatory behavior around of a mean stationary value. The stationary value that each element reaches depends on the value of $\Lambda$. It is important to note that the time during which we followed the evolution is sufficiently large compared with the system time scale defined by $J$. One can appreciate in a qualitative way that, as the number of particles is increased, the deviations from the mean stationary values become smaller as the number of particles is increased. To provide a quantitative estimation of the deviation from the mean stationary value of the elements $\rho_{ij}(t)$  as a function of $N$, we employ the Frobenius norm given by \cite{Meyer} 

\begin{equation}\fl
{\mathcal \sigma}= \left( \sum_{i,j} |\rho_{ij}(t) -\rho_{ij}^s|^2  \right)^{1/2}
\end{equation}
where $\rho_{ij}(t)$ is the value of the matrix element $i,j$ at time $t$ and $\rho_{ij}^s$ is its corresponding mean constant stationary value. This quantity is an extension of the vector norm to a matrix norm and allows us to characterize the fluctuations  from the mean stationary as a function of $N$ \cite{Meyer}. Also we should point out that this quantity does not depend on the basis since it is a trace. We plot ${\mathcal \sigma}$ in the bottom of Fig.\ref{Fig2}.

The second initial condition that we consider belongs also to the SU(3) coherent states. We shall show that following their evolution in the appropriate basis, the characteristic signature of intrinsic decoherence indeed become apparent, namely, that the off diagonal matrix elements become zero on average. We chose $n_1=n_2=n_3=1/3$ and $\phi_{12}=\phi_{23}=\pi$. Because of the symmetry of Hamiltonian \ref{h3} and the fact that this initial condition represents a symmetric state, some of the elements of the reduced one body matrix are equivalent in the Fock basis. 
\begin{figure*}[!htbp]
\centering
{\includegraphics[width=126mm,height=89mm]{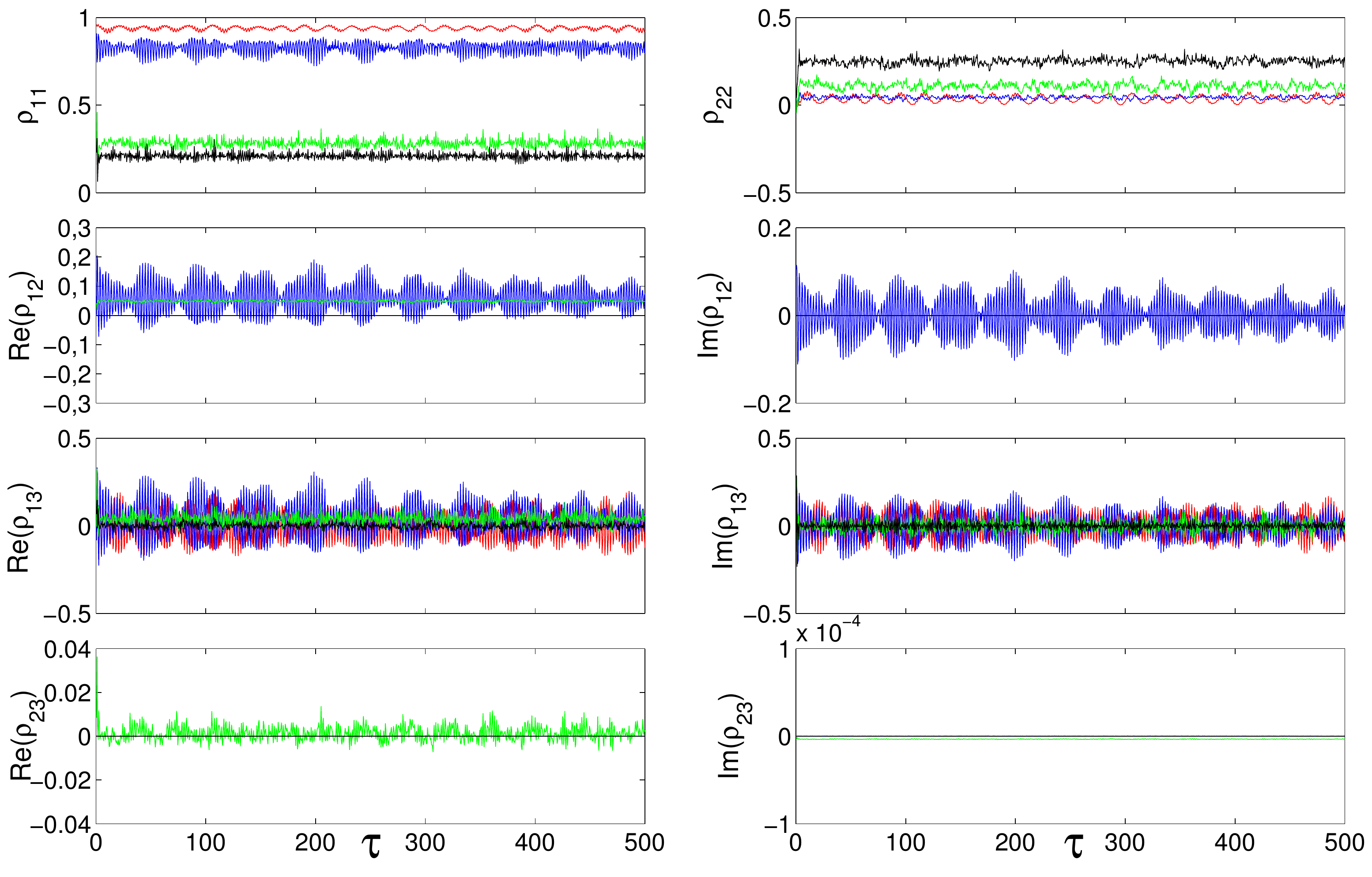}}
{\includegraphics[width=80mm,height=30mm]{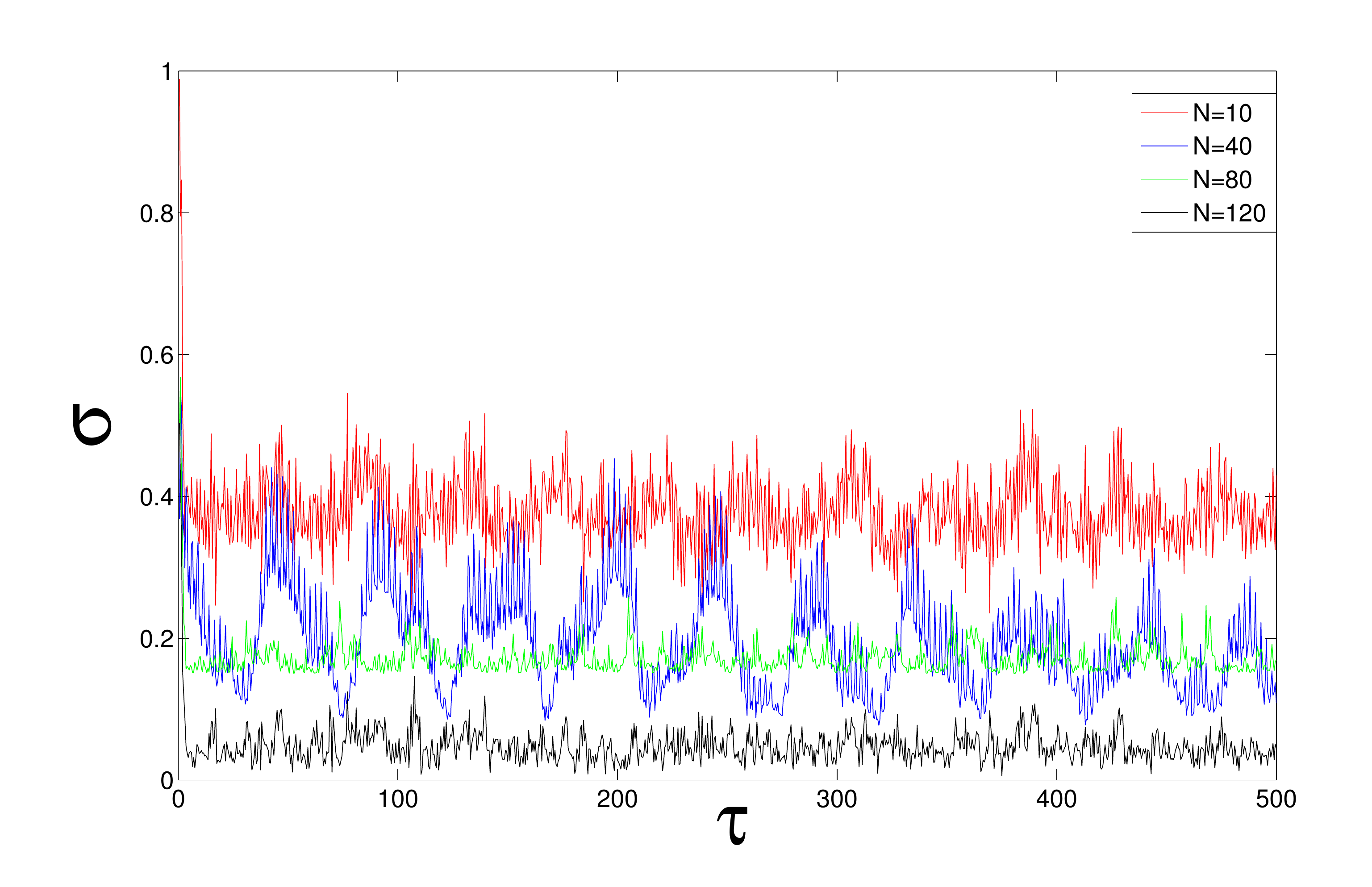}}
\caption{(Color online) Real and imaginary parts of the diagonal and off-diagonal elements of the one-body reduced density matrix as a function of $t$ for $N=10,40,80$ and 120 in the preferred basis. The initial condition is  defined by equation (\ref{SU3}) with $n_1=n_2=n_3=1/3$ and $\phi_{12}=\phi_{23}=\pi$. $\tau$ is in dimensionless units.}
\label{Fig3}
\end{figure*}

As before, we follow the evolution in time of the real and imaginary components of the elements of the one-body reduced density matrix $\rho_{i j}^R$ considering several values of $N$ and keeping constant the value of $U_0$. The chosen initial condition does not correspond to any of the MF stationary solutions of Fig.\ref{Fig1}b, and thus the matrix elements evolve showing an oscillatory behavior. In an analogous way to the initial condition studied above, we identified for each given value of $\Lambda$ the constant mean value that each element $\rho_{ij}$ acquires. With those values we construct the stationary matrix $\rho_s^R$. Then, we numerically determine the unitary transformation defined by $\mathcal{U}$ that allows us to write $\rho_s^R$, that is, the elements of the reduced density matrix in the preferred basis. In Fig. \ref{Fig3} we plot the evolution in time of $\rho^R$ in the preferred basis. As shown in the figure, each of the off-diagonal elements of the reduced density matrix oscillates around zero, showing thus the signature of decoherence in the preferred basis. As in the case of the previous initial condition, one can appreciate the dependence of the fluctuations around the mean stationary value on the number of particles $N$. This information is summarized in panel on the bottom of Fig.\ref{Fig3}.  

\begin{figure*}[!htbp]
\centering
{\includegraphics[width=126mm,height=80mm]{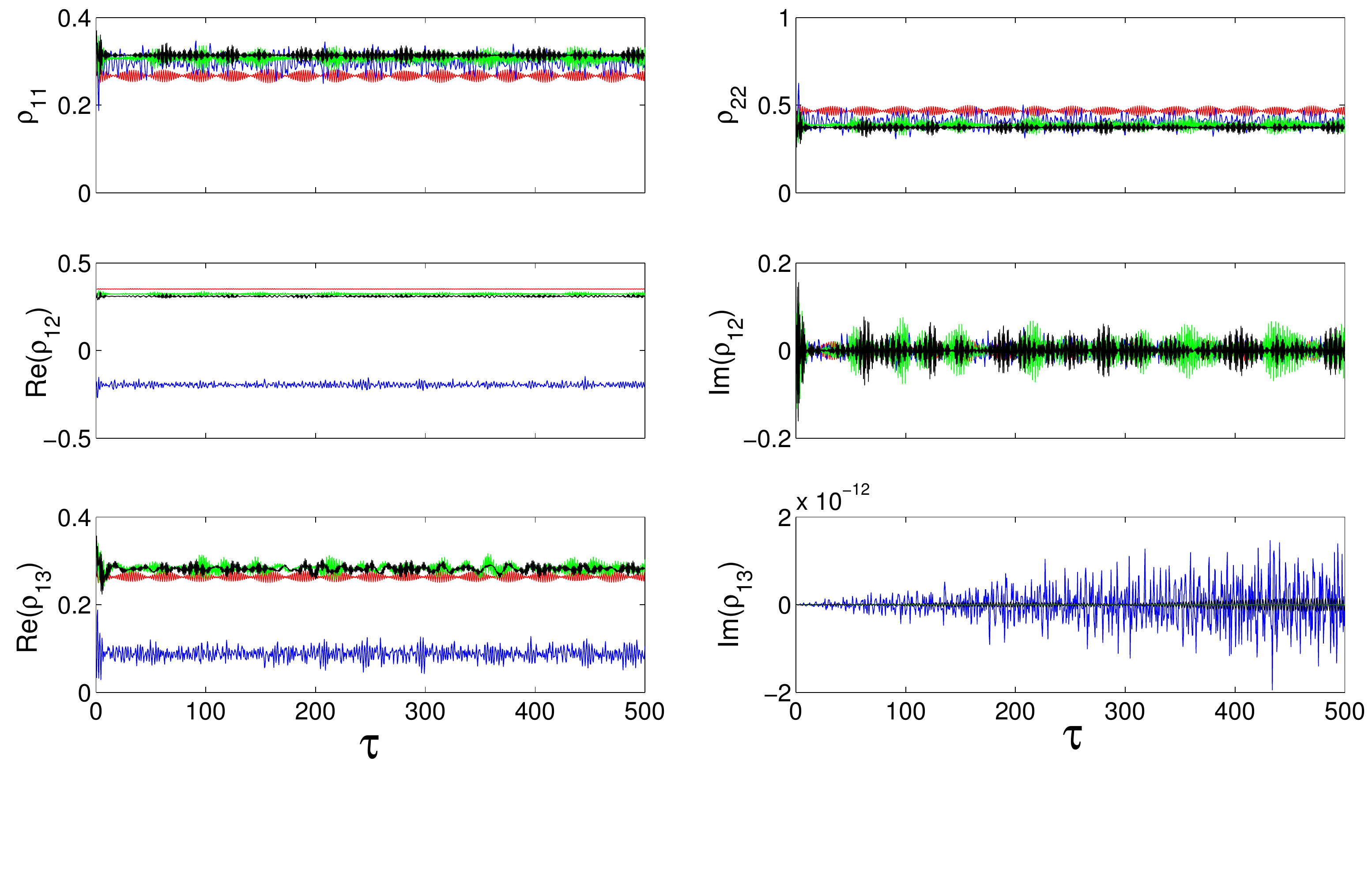}}
{\includegraphics[width=80mm,height=30mm]{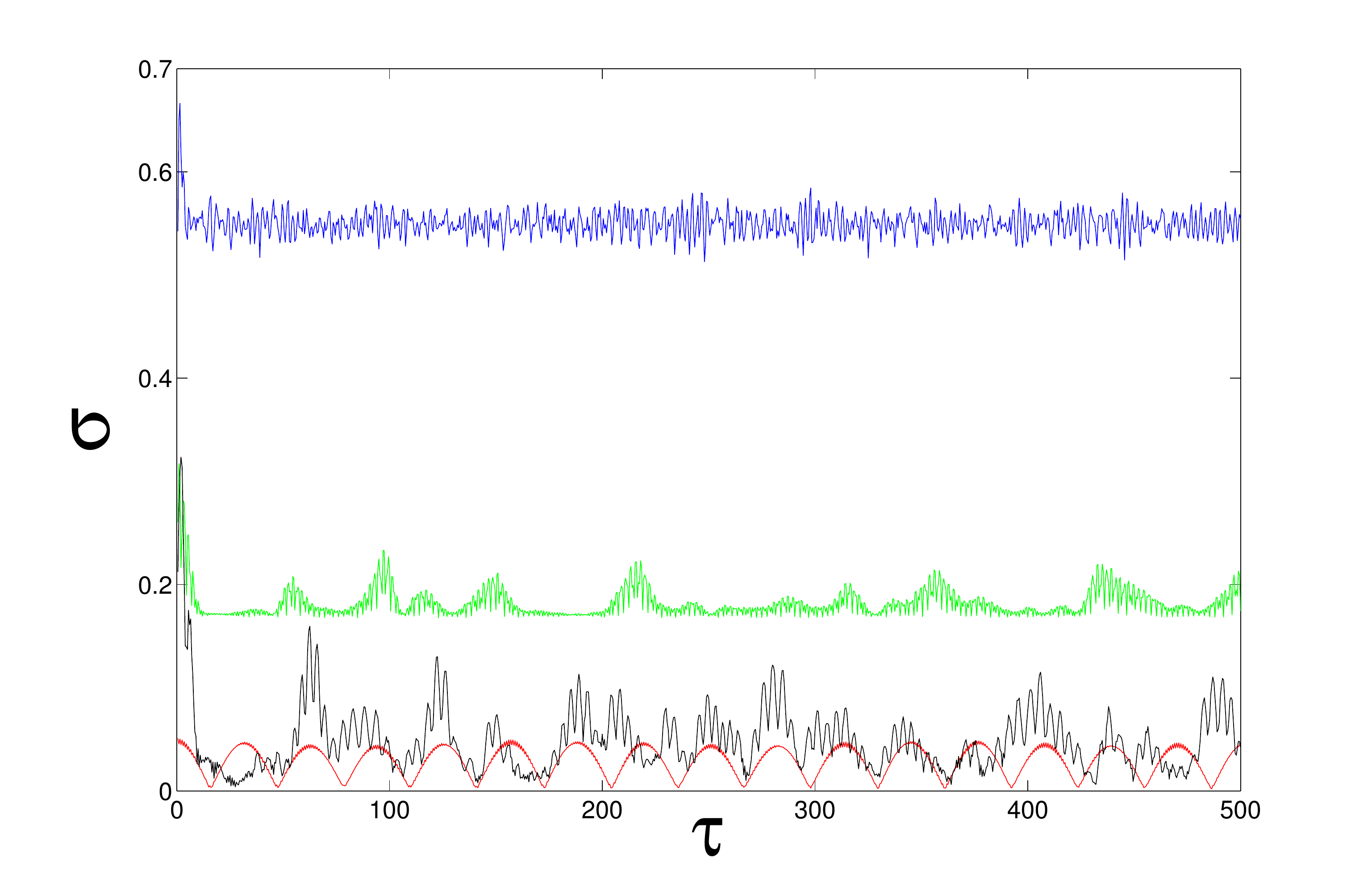}}
\caption{(Color online) Real and imaginary parts of the diagonal and off-diagonal elements of the one-body reduced density matrix as a function of $t$ for $N=10,40,80$ and 120. The initial condition is defined by equation (\ref{SU3}) with $n_1=n_3=0.25$, $n_2=0.5$, and $\phi_{21}=\phi_{32}=0$. $\tau$ is in dimensionless units.}
\label{Fig4}
\end{figure*}

To complement the pictorial catalog that exhibits the existence of the stationary state when the system is evolved from an initial SU(3) coherent state, we selected $n_1=n_3=0.25$, $n_2=0.5$, and $\phi_{21}=\phi_{32}=0$. These values of the conjugate variables $n_i$ and $\phi_{ij}$ for $\Lambda =0$ correspond to one of the MF stationary solutions, in the phase diagram this fixed point is the first of the red lower branch of Fig.\ref{FigA}. In Fig.\ref{Fig4} we plot the real an imaginary components of the one body reduced density matrix in the Fock basis. We notice that for every value of $N$ the amplitude of the oscillations is smaller than that associated to the previous initial conditions (see Figs. \ref{Fig2} and \ref{Fig3}), that is, correspond to small deviations from the stationary state. In particular, for $N=10$ (red color), $\Lambda =0.5$ and thus the oscillations of each element of the reduced density matrix are very small. As in the previous analysis we include a plot of ${\mathcal \sigma}$ to quantify the deviations from the mean stationary value. 

\begin{figure*}[!htbp]
\centering
{\includegraphics[width=126mm,height=80mm]{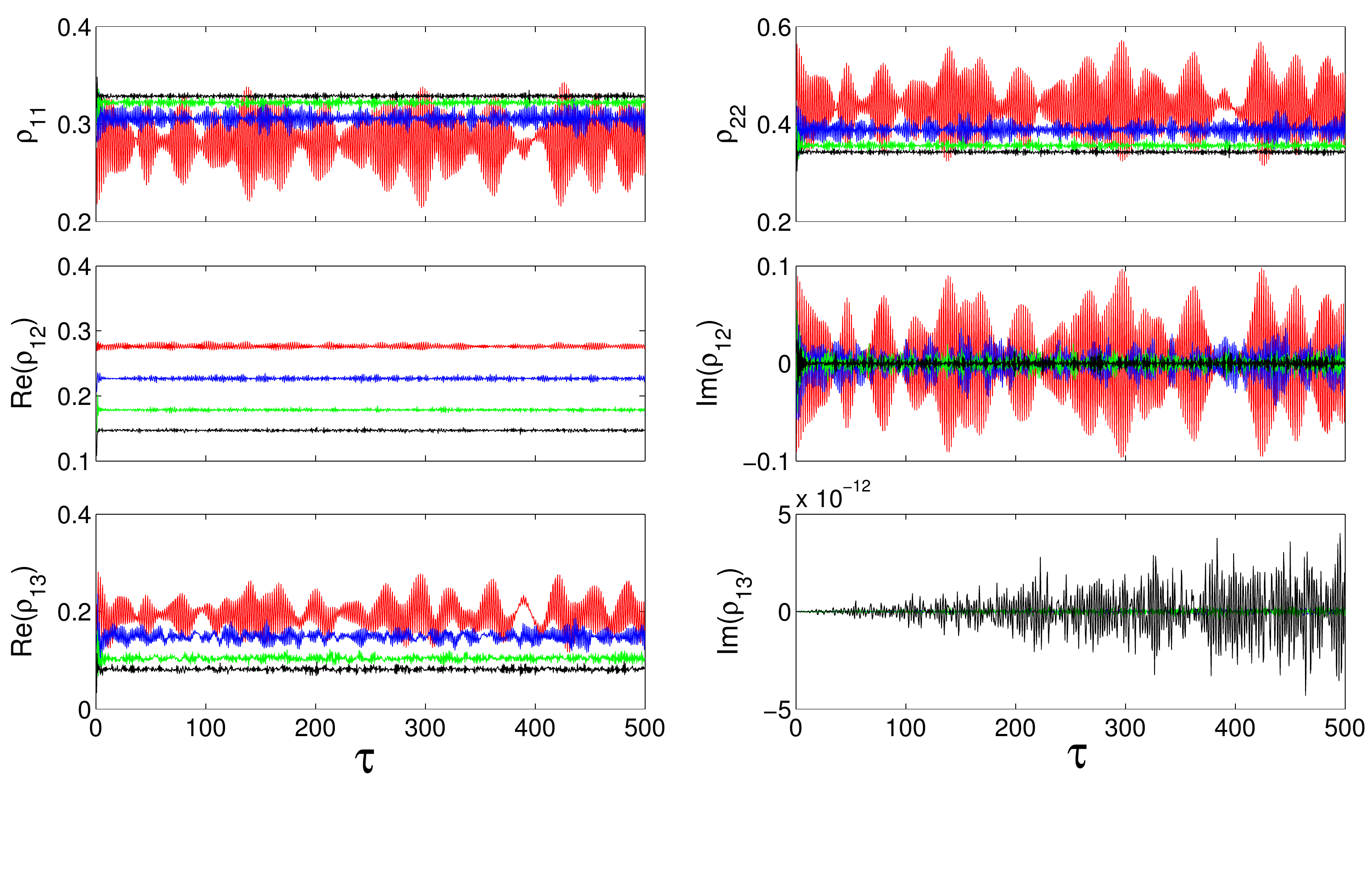}}
{\includegraphics[width=110mm,height=40mm]{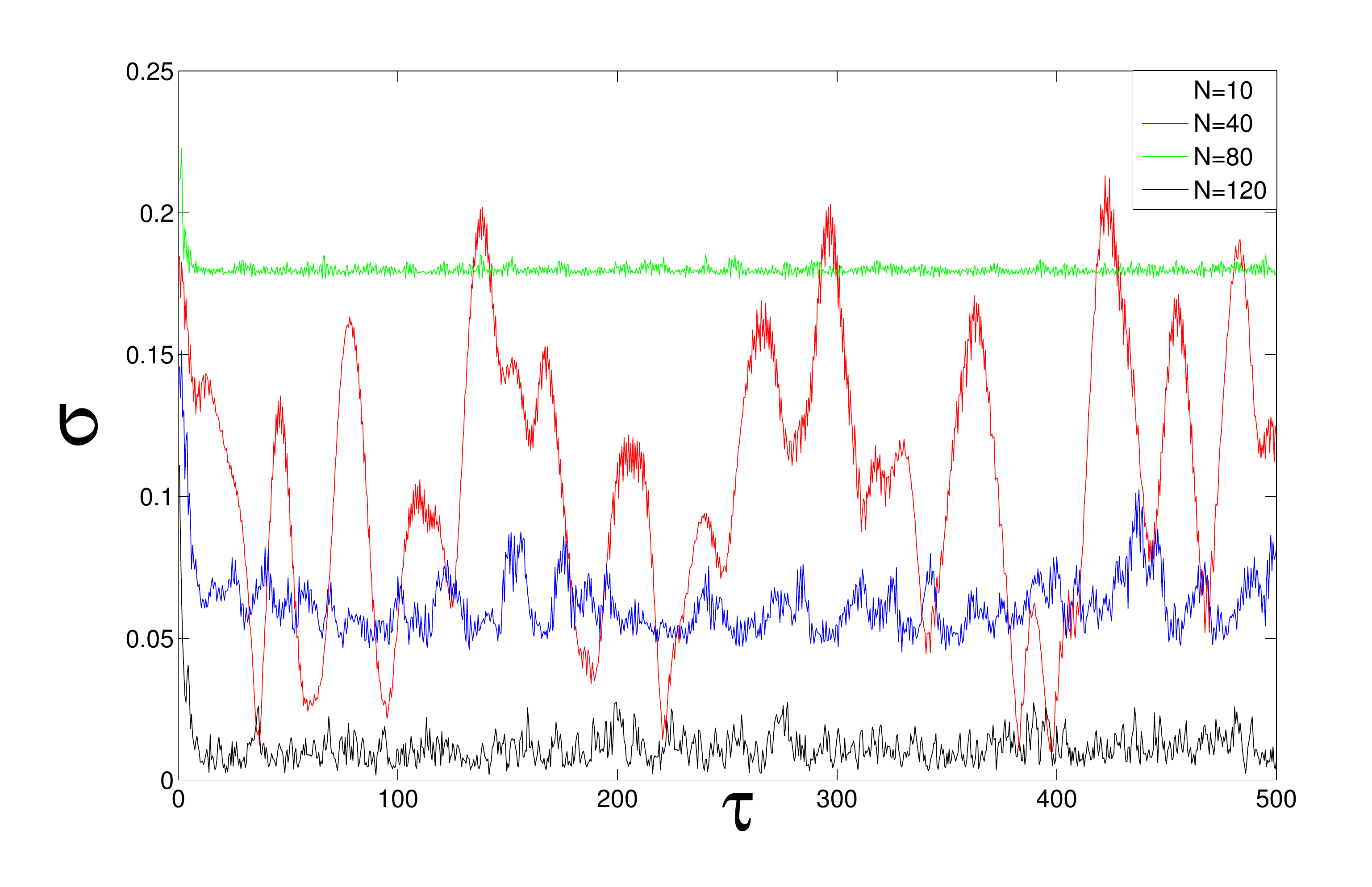}}
\caption{(Color online) Real and imaginary parts of the diagonal and off-diagonal elements of the one-body reduced density matrix as a function of $t$ for $N=10,40,80$ and 120. The initial condition is $|\phi(0) \rangle = 1/\sqrt{\Omega} \sum _i^\Omega |i\rangle$. $\tau$ is in dimensionless units.}
\label{Fig5}
\end{figure*}

Regarding the number Fock states $|n_1,n_2,n_3 \rangle$, which as a matter of fact are eigenstates of the interaction term of Hamiltonian (\ref{h3}), one observes that when the system is evolved from each basis element, a behavior similar to that associated with the SU(3) family of coherent states is found, namely, that the time evolution of $\rho_{ij}$ show oscillations with a maximum amplitude, then becoming damped until the mean stationary state is reached. To conclude the study of the intrinsic decoherence we select as initial state $|\phi(0) \rangle = 1/\sqrt{\Omega} \sum _i^\Omega |i\rangle$, where $i$ labels the Fock states, $|1\rangle = |N,0,0 \rangle$, $|2\rangle = |N-1,1,0 \rangle$, $|3\rangle = |N-1,0,1 \rangle$ and so on, and follow its time evolution for several values of $\Lambda $ in the interval $0.5 <\Lambda<6$. It is worth to mention that this state cannot be written as a SU(3) state.  In Fig. \ref{Fig5} we plot real and imaginary components of $\rho^R$ in the Fock basis (upper panel) and ${\mathcal \sigma}$ in the lower panel. As one can see from these figures, one can confirm again the existence of the statistically stationary state. 

As it is well known the intrinsic difference among MF and BH treatments is that while in MF the transition from JO to ST regimes is promoted by the nonlinear term that scales with the parameter $\Lambda$, in BH such transition caused by the interaction among the particles. In the same way, our calculations showed that the stationary state is also a consequence of dealing with a many body description. To verify this statement we consider values of $\Lambda$ below and above the transition in both MF and BH schemes for the analyzed initial conditions. We found, as expected, that the Frobenius norm is always large in MF than in BH as the number of particles increases.
 
\subsection{Purity}
 
 Another dynamical aspect that we can study within the BH scheme and that also emerge as a consequence of the presence of particle-particle interactions is the degree of mixing that a state has.  The purity of an $N$-particle system is defined as
\begin{eqnarray}\fl \nonumber
\mathcal{P}(t)={\rm Tr} (\rho^N(t)^2)={\rm Tr} (U^\dagger\rho^N_0U U^\dagger \rho^N_0 U)\\ ={\rm Tr}( (\rho^N_0)^2),
\label{purity}
\end{eqnarray}
 where $U$ is the evolution operator defined in section I, and $\rho^N_0 $ represents the $N-$body density matrix at $t=0$. As referred above, such quantity is an effective measure that specifies the degree of mixing of a state \cite{Viscondi1,Chianca, Somma}. A pure or coherent state is characterized by having $\mathcal{P}=1$ while $\mathcal{P} < 1$ indicates a that the state is mixed. If the initial state is a pure, which means that a complete set of measurements has been performed, it remains pure for all times, whenever such initial state be evolved in time under the action of the unitary operator $U$. However, an interesting question to be addressed is the investigation of the purity of few-body states, when they are initially described by a pure state. By making the analogy with the intrinsic decoherence extracted from the one-body reduced density matrix, we investigate here the purity of the reduced density matrix $\rho^R$ to analyze the role of the interactions in transforming an initial pure state into a mixed one.  

We define the purity $\mathcal{P}$ of the one-body reduced density matrix $\rho^R$ as 
\begin{equation}\fl
{\cal P}={\rm Tr}  (\rho^R(t)^2).
\end{equation}
where
\begin{equation*}\fl \footnotesize
 {\rm Tr} ((\rho^R)^2)=\sum_{i,j}\langle\phi(t)|b^\dagger_jb_i|\phi(t)\rangle\langle\phi(t)|b^\dagger_ib_j|\phi(t)\rangle,\end{equation*} being $|\phi(t)\rangle$ an $N-$body system state, that is, a pure state. By taking the derivative of $\mathcal{P}(t)$ with respect of time $t$ one finds
\begin{eqnarray}\fl\nonumber
i\hbar\frac{\partial {\rm Tr} ((\rho^R)^2)}{\partial t}=\\ \nonumber \fl2\sum_{i,j}\langle\phi(t)|[b^\dagger_jb_i,H_0+H_I]|\phi(t)\rangle\langle\phi(t)|b^\dagger_ib_j|\phi(t)\rangle,
\end{eqnarray}
where $H_0=- J(b_1^{\dagger}b_{2}+b_2^{\dagger}b_{1}+b_2^{\dagger}b_{3}+b_3^{\dagger}b_{2})$ and $H_I =U_0\sum_{i=1}^3b_i^{\dagger}b_i^{\dagger}b_{i}b_{i}$. From the usual commutation relations for bosons it follows that 
$$\sum_{i,j}\langle\phi(t)|[b^\dagger_jb_i,H_0]|\phi(t)\rangle\langle\phi(t)|b^\dagger_ib_j|\phi(t)\rangle=0.$$
By identifying $\alpha_{i,j}=\langle\phi(t)|b^{\dagger}_i b_j|\phi(t)\rangle$, $\beta_{i,j,k,l}=\langle\phi(t)|b_i^{\dagger}b_j^{\dagger}b_kb_l|\phi(t)\rangle$ and rescaling the time $t$ to dimensionless units  $\tau= Jt/\hbar$, one can write the final expression for the time derivative of the purity of the reduced density matrix,

\begin{eqnarray}\fl\nonumber
\frac{\partial {\rm Tr} ((\rho^R)^2)}{\partial \tau}=-4\Lambda\sum_{i\neq j}{\rm Im}\left(\alpha_{ij}(\beta_{jiii}-\beta_{jjji})\right).\\
\label{purityDt}
\end{eqnarray} 
Since in general the right hand side of the last equation is different of zero, one concludes that when an initial pure state is evolved in time it becomes a mixed one. We should notice however that the right hand side becomes zero when the stationary state has been reached.

\begin{figure*}[!htbp]
\centering
  \begin{tabular}{cc}
{\includegraphics[width=80mm,height=40mm]{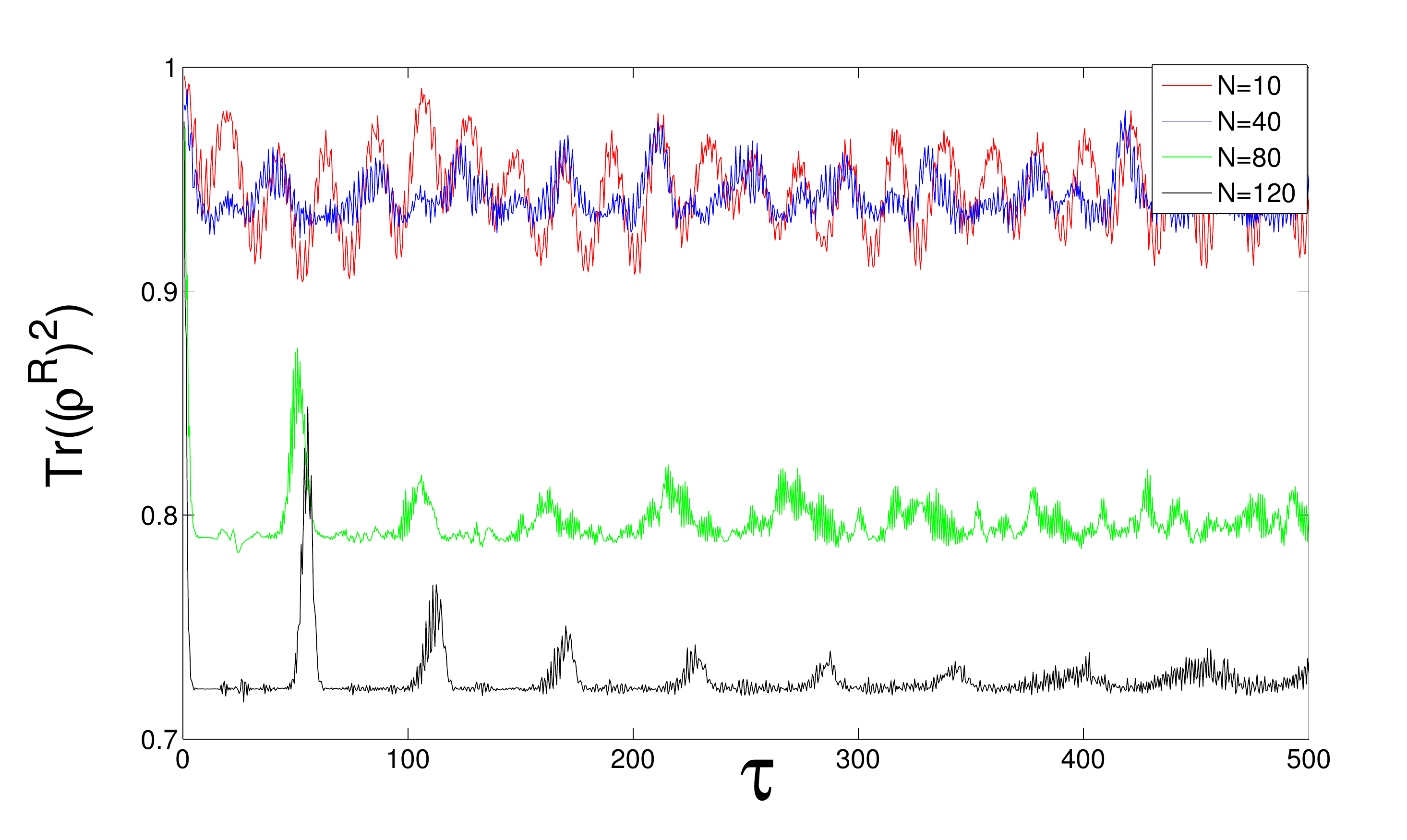}}&
{\includegraphics[width=80mm,height=40mm]{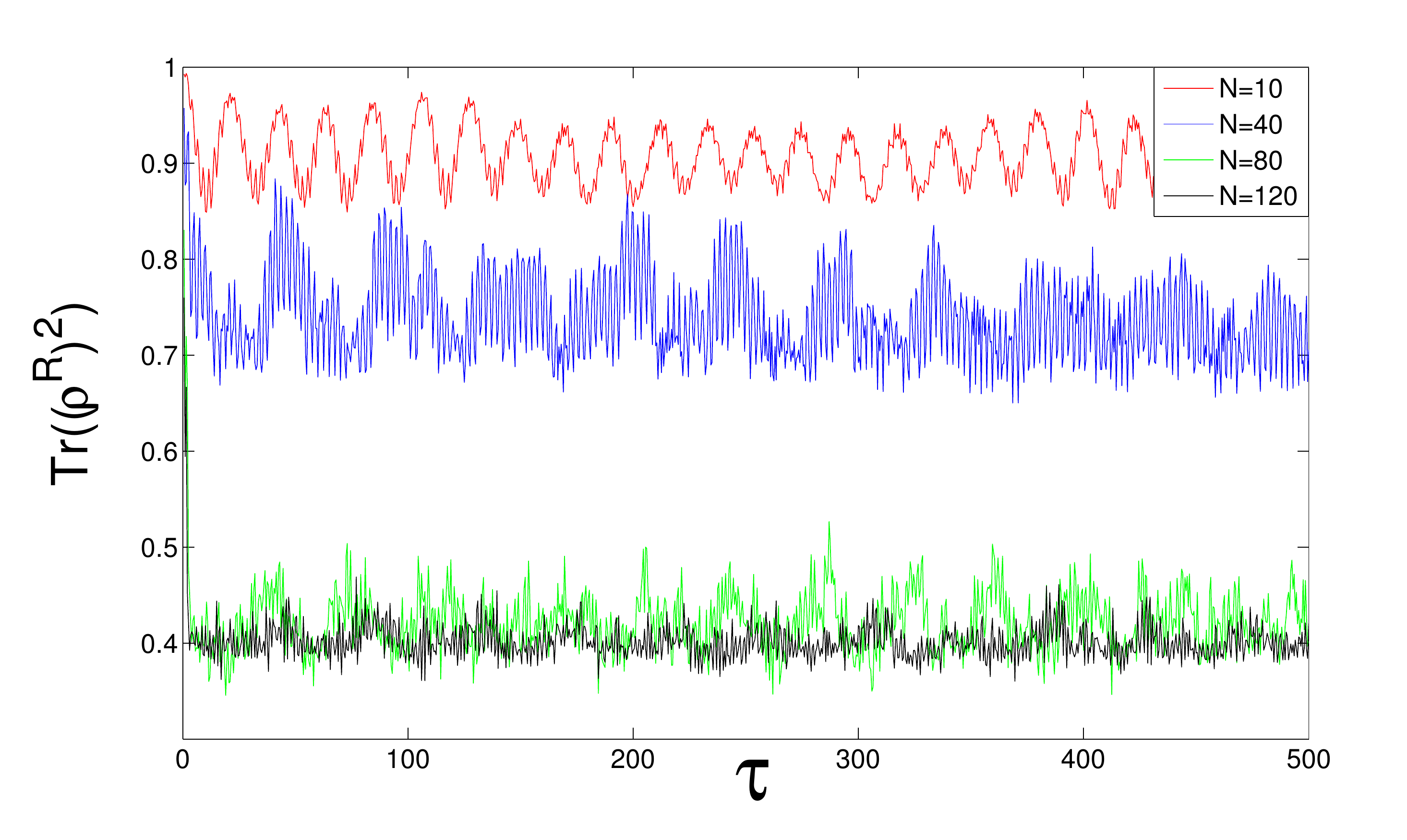}}\\
{\includegraphics[width=80mm,height=40mm]{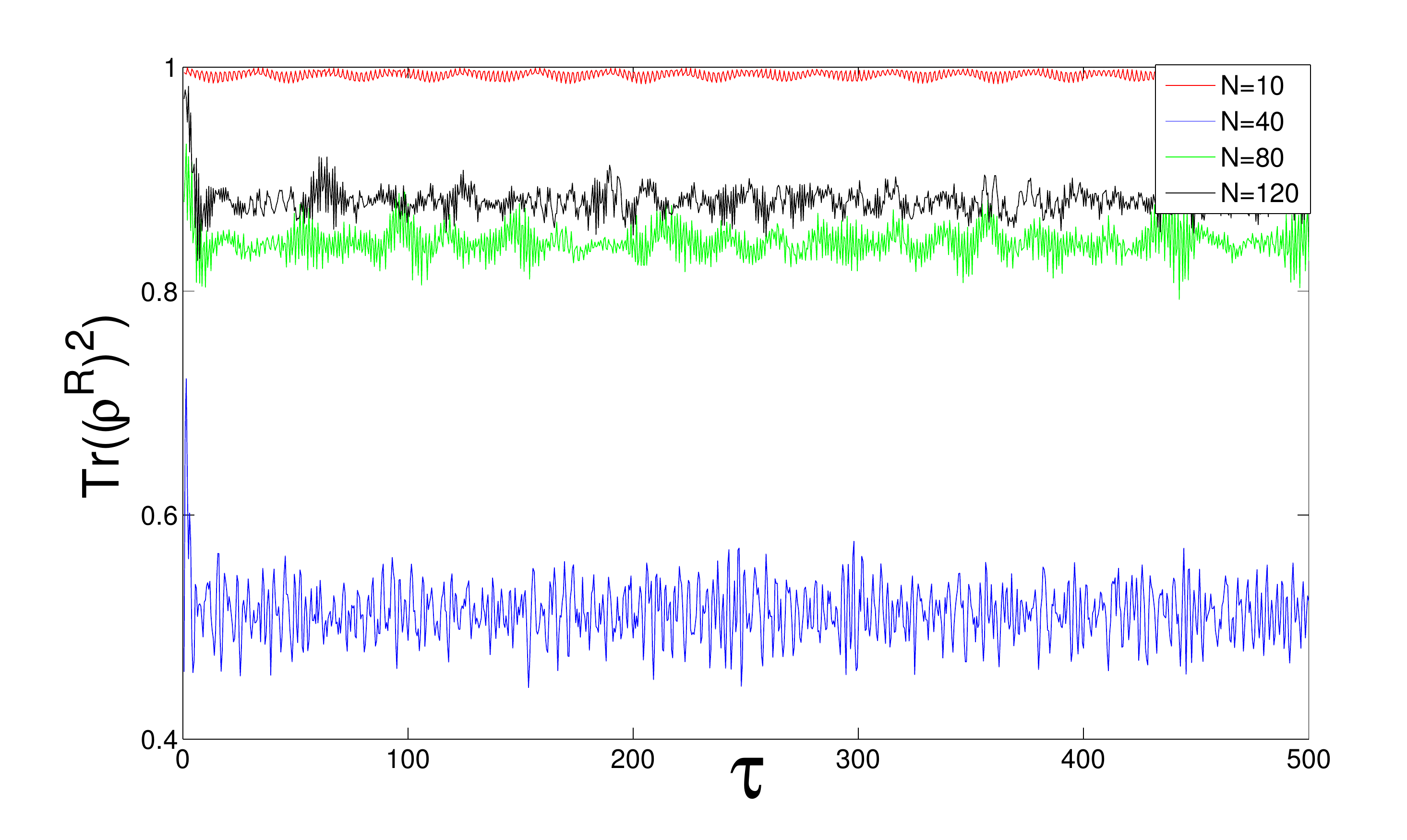}}&
{\includegraphics[width=80mm,height=40mm]{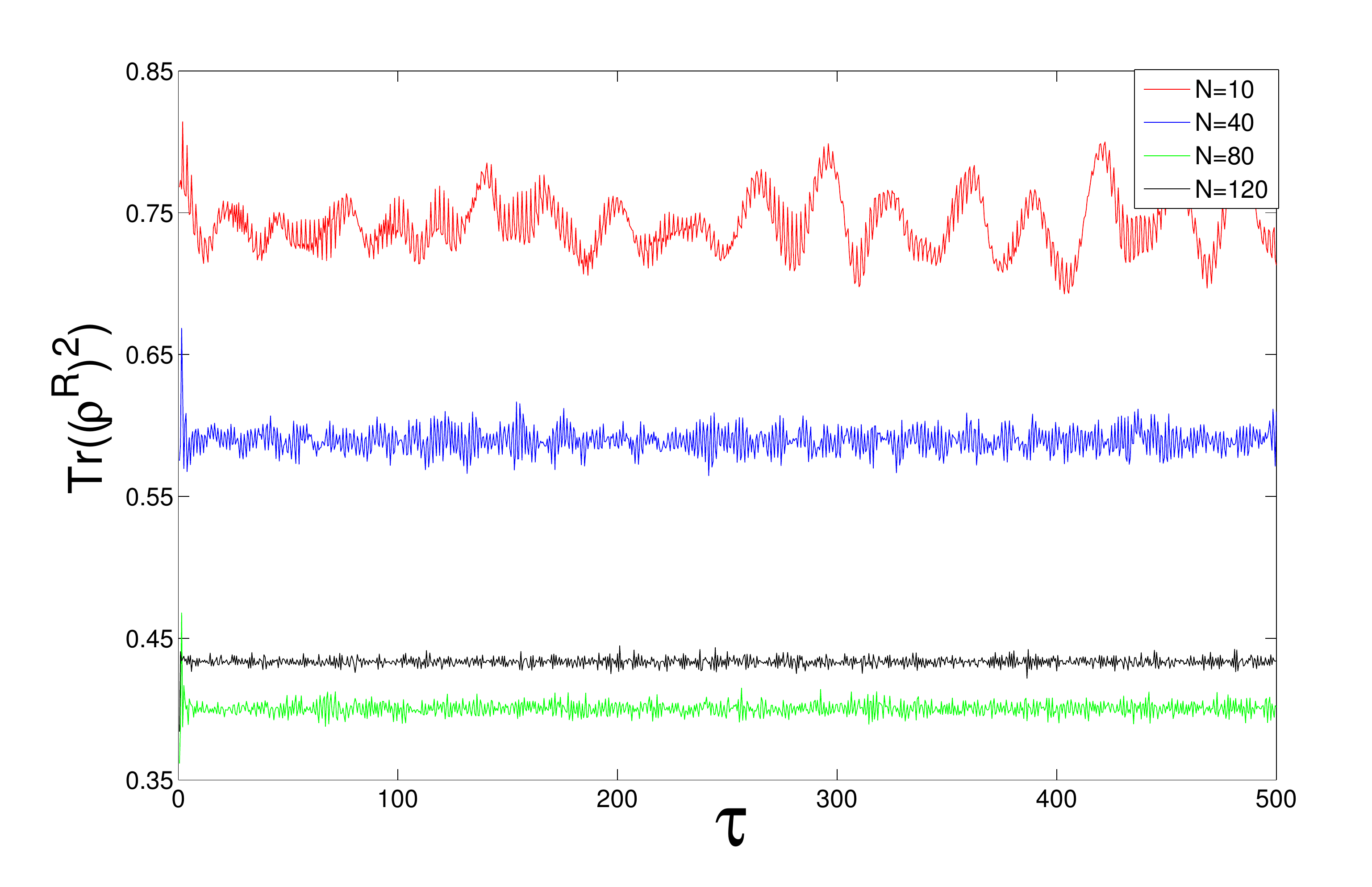}}
\end{tabular}
\caption{(Color online) Purity as a function of time for the initial conditions of Figs. \ref{Fig2} (first row, left), \ref{Fig3} (first row, right), \ref{Fig4} (second row, left) and \ref{Fig5} (second row,left). $\tau$ is in dimensionless units.}
 \label{Fig6}
\end{figure*}

To illustrate the behavior of the evolution in time of the purity for the one-body density matrix we chose as initial states those considered above for the analysis of the intrinsic decoherence. Captions in figures \ref{Fig2}, \ref{Fig3}, \ref{Fig4} and \ref{Fig5} specifies these initial conditions. All of these states have an initial purity equal to 1. The considered values of $\Lambda$ are indicated in Fig. \ref{Fig6}. 

From Fig. \ref{Fig6} one can confirm, as predicted from Eq. (\ref{purityDt}) that, for any arbitrarily small value of $\Lambda$, the purity decreases, thus showing the tendency of an initial state to become mixed. In this case we also appreciate the role of the system size, particularly, when the state evolves towards a mixed state. The information provided by the purity is complementary to the information obtained from ${\mathcal \sigma}$. We observe for example that in the case in which the initial state is very close to an stationary MF solution (see red color on the third row of Fig. \ref{Fig6}) the state remains essentially pure along the time evolution. It is important to notice that the measure of the purity does not depend of the basis. 

In the lower panel of Fig. \ref{Fig6} it is shown the purity for an initial state given as a superposition of Fock states $|\phi(0) \rangle = 1/\sqrt{\Omega} \sum _i^\Omega |i\rangle$,  for several values of $N$. One observes from this figure that for a given value of $\Lambda$ the purity remains essentially constant. From the analysis of the purity for the initial conditions in Fig. \ref{Fig6}, and the present one, we conclude that for any arbitrary initial state is impossible to increase the coherence. The time window in which the coherence is apparently increased is small compared with the time in which it diminishes.

\section{Final Remarks}
We have investigated the dynamics and the onset of stationarity in an interacting BEC confined in a triple well potential in 1D. The analysis was performed in the Bose-Hubbard framework by studying the evolution in time of one-body properties. In particular, we concentrated in examine the role of the system size. In our study we revisited the analysis of the stationary states and distinguished among three kinds of stationary states. The semiclassical stationary states predicted within the MF frame, the truly stationary states or eigenenergy states obtained by direct diagonalization of the Hamiltonian written in the BH scheme, and the statistically stationary states revealed from the evolution in time of the one-body reduced density matrix. 

On the dynamical side we studied two aspects that can only be accounted within the BH scheme, namely, effects of intrinsic decoherence and the mixing of an initial pure state. To perform such an analysis we considered the evolution in time of one-body properties in the $N-$particle environment. From the time behavior of the one-body reduced density matrix we reached the conclusion that within fluctuations around a mean constant value, stationarity is always observed in every closed BH-like. To determine the role of the system size in the observation of an stationary state we quantified the deviations from the mean stationary value in terms of a matrix norm, the Frobenius norm. We studied the evolution of the one-body reduced density matrix to trace for signatures of decoherence, namely, we show that it is always possible to follow the evolution in a particular basis where the off-diagonal elements of the reduced density matrix become, within the fluctuations, zero in the stationary state. Regarding the purity we also considered the one-body reduced density matrix to show the influence of the particle-particle interactions in transforming an initial pure state into a mixed one. We demonstrated that this conclusion remains valid for Hamiltonians with Bose-Hubbard-like structure. Our numerical calculations for $N=10, 40, 80, 120$ and 150 allowed us to verify that the onset of stationarity is also consequence of dealing with a many body description and not an effective result driven by a non-linearity as the transition among JO and ST in the MF approach. 

According to the fundamental hypothesis of statistical physics, the existence of a stationary state is based on the observation that every isolated many-body system, whose energy is sharply defined, and that is left unperturbed, attains a state that no longer evolves in time, the equilibrium state. This state is detected when few-body physical quantities are measured, as for example, density, temperature, pressure, energy and compressibilities among others. We believe that the statistically stationary states studied here are analogs of equilibrium states. We do such an analogy since the Bose atoms confined in the triple well potential (and in general an optical lattice) constitute a closed system in which the dynamics of arbitrary initial states is dictated by the evolution operator $U= e^{-i H t/\hbar}$, that leads the system towards a stationary state. It thus appears that intrinsic decoherence is a synonymous of relaxation towards an equilibrium state in a closed system, and that the typical observation of equilibrium thermodynamic states in closed systems, are a consequence of intrinsic decoherence.

\ack

This work was partially funded by grants IN108812-2 DGAPA (UNAM) and CB09-132527 (CONACYT). A.C.G  acknowledges support from CONACYT.

\appendix

\section{}

\begin{figure*}[htbp]\centering
\includegraphics[width=3.0in]{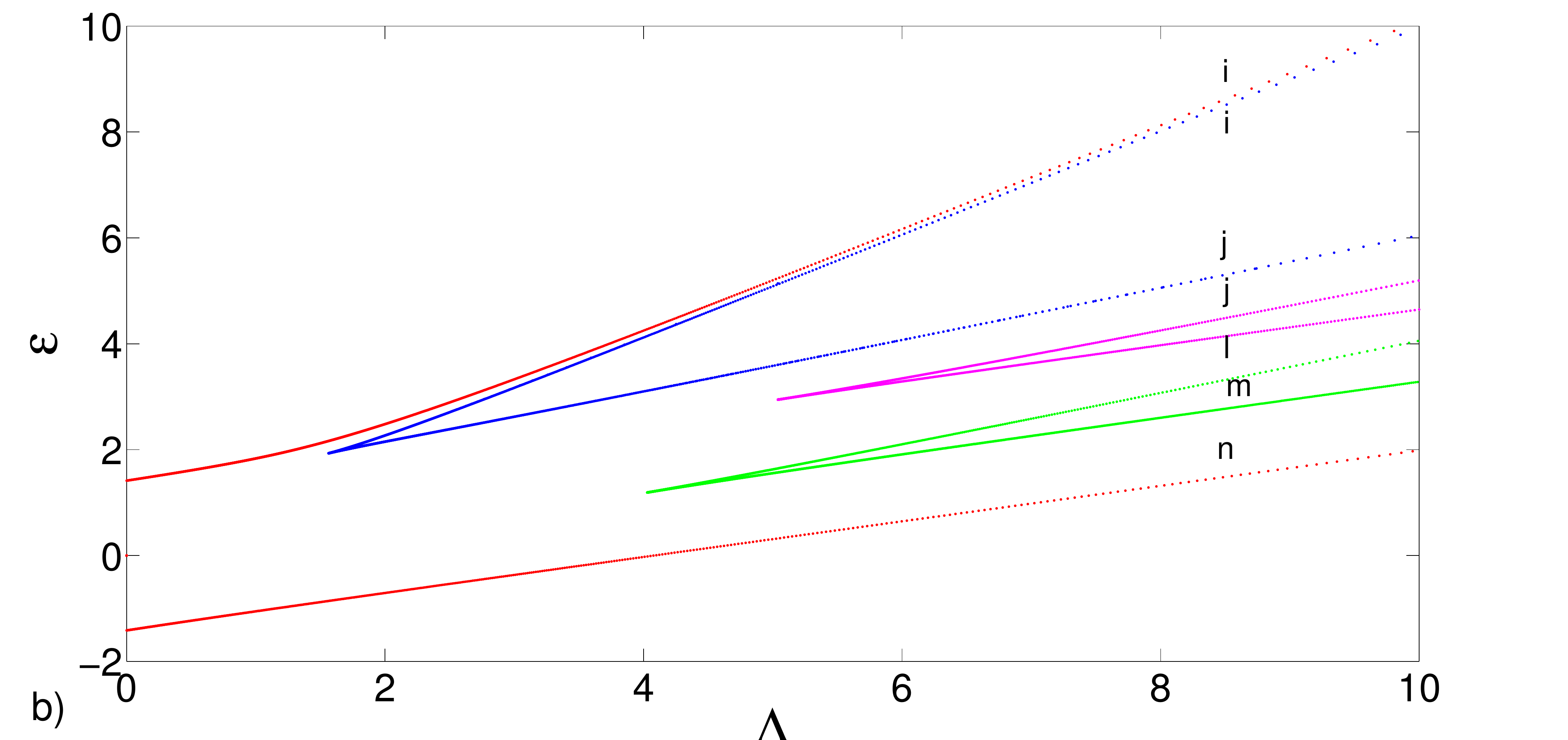}
\caption{(Color online) Stationary energies as a function of $\Lambda =U_0N/J$ for MF scheme. Letters on the curves indicate the number of negative eigenvalues of the Hessian matrix $ \left( \frac{\partial^2 H }{\partial x_i \partial x_j} \right )$, i,  j, l, m, n correspond to 4, 3, 2, 1 and 0 respectively. }
\label{FigA}
\end{figure*}

Let us to start by considering the Gross-Pitaevskii equation for the macroscopic wave function $\Psi(x,t)$, which in the case of a triple well potential represented by $V_{ext}(x)$, adopts the form
\begin{eqnarray*}\fl
i\hbar\frac{\partial \Psi(x,t)}{\partial t}=\left[-\frac{\hbar^2}{2m}\nabla^2
+V_{ext}(x)\right]\Psi(x,t)\\+g|\Psi(x,t)|^2 \Psi(x,t).
\end{eqnarray*}
In terms of the approximate three lowest eigenfunctions of the single-particle Hamiltonian $H_0= -\frac{\hbar^2}{2m}\nabla^2+V_{ext}(x)$ \cite{Paredes}, one can write the localized basis states $\psi_j(x)$, $j=1,2,3$ and then substitute it into the macroscopic wave function $\Psi(x,t)=\psi_1(x)\Phi_1(t)+\psi_2(x)\Phi_2(t)+\psi_3(x)\Phi_3(t)$, being the time dependence included in $\Phi_j(t)$. Thus, by neglecting constant energy factors proportional to $E_j= \int \left(-\frac{\hbar^2}{m} |\nabla \psi_j|^2 + |\psi_j |^2 V_{ext} \right)dx$ we arrive at the equations $ i\hbar\frac{\partial \Phi_j(t)}{\partial t}=U_j\,N_j\Phi_j-J_{j2}\Phi_2$ and  $i\hbar\frac{\partial \Phi_2(t)}{\partial t}=U_2\,N_2\Phi_2-J_{12}\Phi_1-J_{23}\Phi_3$ where $j=1,3$,
$U_j= g\int |\psi_j|^4dx$, $J_{12}=-\int \left( \nabla \psi_1 \cdot \nabla \psi_2 + \psi_1 V_{ext} \psi_2\right)dx $, $J_{23}=-\int \left( \nabla \psi_2 \cdot \nabla \psi_3 + \psi_2 V_{ext} \psi_3\right) dx$ and  $N_i$, $i =1,2,3$, represent the particle population in wells left, center and right respectively. The stationary states of the non-linear Schr\"odinger equation are $\Phi_j(t)=\sqrt{N_j}\exp i(\phi_j-\nu t)$. Thus, by assuming $U_i=U_0$ and $J_{ij}=J$ for $i,j=1,2,3$, and substituting $n_i=N_i/N$ we find

\begin{eqnarray}\fl \nonumber
\varepsilon=- 2\sqrt{n_1n_2} \cos{\phi_{21}} -2\sqrt{n_2n_3} \cos{\phi_{23}}\\
+\Lambda\left(n_1^2 +n_2^2+n_3^2 \right) ,
\label{Hsc}
\end{eqnarray}
where $\varepsilon$ is the dimensionless energy per particle $\varepsilon=\hbar \nu/NJ$, and $\phi_{ij}=\phi_j-\phi_i$ measures the phase difference among the condensates in wells $i$ and $j$. One can see that $\varepsilon$ is a function of $n_i$ and $\phi_{ij}$ and, thus, we require to explore such a phase space to determine the values that produce a solution $\varepsilon$ for a given value of $\Lambda$. This can be done by associating a classical Hamiltonian the constant energy $\varepsilon$, and then imposing stationarity to the corresponding Hamilton equations for the conjugate variables $(\phi_{ij},n_i)$. Such stationary solutions correspond to maximum, minimum or saddle points. The stability character of each solution was determined by means of the eigenvalues of the Hessian matrix $\left( \frac{\partial^2 H }{\partial x_i \partial x_j} \right )$, being $x_i$ the phase difference $\phi_{ij}$ or the population $n_i$. From the eigenvalues of this matrix we established if a given stationary solution corresponds to a maximum, minimum or saddle point. We recall that these extrema correspond to having all the eigenvalues negatives, all the eigenvalues positive or eigenvalues with different signs, respectively. In Fig. \ref{FigA} we summarize all the stationary states consistent with the Hamilton equations for $n_i$ and $\phi_{ij}$. In the same figure we indicate the number of negative eigenvalues of the Hessian matrix associated to each stationary solution to distinguish the stability character of each solution. 
We observe from Fig. \ref{Fig1} that there is a discrepancy in the number of stationary states predicted by each approach. We attribute such lack of compatibility to the essential difference between the two approaches, namely that MF treatment is always an effective equation for one-body properties, while the BH approach considers one- and two-body terms from which effects of $m$-body properties (with $m<N$) can be tracked down. In particular, the expectation value of the Hamiltonian in the eigenstate basis, the energy spectrum, which is a two-body property. This allows us to stress that within the MF approach, in general, neither stationary properties nor dynamical ones involving two or more bodies can be extracted. On the other hand, the inclusion of particle-particle collisions within the BH approach does permit probing any arbitrary $m-$particle property, either stationary or dynamical. 
\end{multicols}

\end{document}